\documentclass[aps,pra,epsfig]{revtex4}

\usepackage[dvips]{graphicx}

\usepackage{epsfig}
\usepackage{soul}
\begin{document}

\title{Inevitability of gauge potentials and the Lagrangian}
\date{\today}

\author{A. R. P. Rau\footnote{arau@phys.lsu.edu}}
\affiliation{$^1$Department of Physics and Astronomy, Louisiana State University, Baton Rouge, Louisiana 70803, USA}

\begin{abstract}
In seeking a minimal variational formulation of Maxwell's equations, one is led naturally to the scalar and vector potentials as ``adjoint" functions in a well-defined sense and to the crucial minus sign that defines the Lagrangian.

\end{abstract}

%\pacs{03.65.Ta,03.67.-a}

\maketitle

After its beginnings with Galileo and Newton, surely one of the most important developments in physics was the Euler-Lagrange-Hamilton formulation of the equations of motion of classical mechanics. Its extensions later to electromagnetism, to General Relativity, and then to quantum mechanics and quantum field theories has made this Lagrangian approach the preferred one throughout physics. The ability to handle generalized coordinates and not just Cartesian, to use scalar energies and potentials rather than vector forces, to be cast as a variational or stationary principle, to handle constraints of motion (through Lagrange multipliers), and the immediate connection to the conservation laws of physics such as of energy, momentum, angular momentum, and charge all make this approach the one of choice, and one introduced early to physics students. 

Two motivations led to this essay. The first is one encountered by generations of physics students including that of the author and his classmates upon first introduction of this object called the Lagrangian with its difference between kinetic ($T$) and potential ($V$) energies. The sum has a natural meaning as the total energy of the physical system and has associated one of the most fundamental laws, of the conservation of energy, but what is the physical significance of taking another combination, also with dimensions of energy, but the difference between the two? The same applies to electromagnetism, where $\vec{E}^2 + \vec{B}^2$ to within numerical factors is the energy density of electric and magnetic fields, but the Lagrangian for those fields involves again the difference \cite{ref1}. Why that sign change and what is its physical significance or origin are questions successive generations seem to continue to wrestle with. The second motivation arose from a remarkable essay many years ago by Dyson with the title ``Feynman's proof of the Maxwell equations" \cite{ref2}. In it, he presents an argument given to him by Feynman in 1948, one that was never published, that attempts to use Newton's laws of motion and the commutation relation between position and velocity to obtain Maxwell's equations. Feynman was searching for something new and beyond standard electrodynamics and, in not finding it, had dropped the subject. An aspect of the Dyson essay that also occurs in many places, namely the treatment of Maxwell's equations as two pairs, motivated the study below. In their usual and familiar form, the two pairs are made of a vector and a scalar equation each, that symmetry in number and form itself a key motivation for what follows.

Before proceeding to the main content of this paper, since the Dyson essay generated many discussions and explorations along various lines, much of that literature having appeared in this journal, we consider some of them briefly in this paragraph. Hughes \cite{ref3} followed more generally Dyson's approach of using non-relativistic classical dynamics and the usual Lagrangian form of $T-V$, while Hojman and Shapley \cite{ref4} investigated the existence of a Lagrangian starting from equations of motion and a very general commutation relation between position and velocity. Vaidya and Farina \cite{ref5} concluded that Maxwell's equations and Galilean mechanics ``cannot coexist peacefully," also disputing Dyson's treating one pair of Maxwell's equations as ``merely" defining charge and current density. However, a consistent ``Galilean electromagnetism" has been discussed \cite{ref6}, along with the existence of two vector fields but with no claim that these are the electric and magnetic fields as usually understood. Heras investigated whether Maxwell's equations can be obtained from the continuity equation, assuming a scalar (charge) and a vector (current) density, arguing for two fields and four coupled equations between them, charge conservation seen as a key element \cite{ref7}. Yet another approach of starting from the Lorentz force, together with space and time reversal invariance to derive Maxwell's equations \cite{ref8} has also a good summary of the various systems of units used, including the Heaviside-Lorentz ones we use below for the equations. We refer the reader to these papers, references therein and in a general review \cite{ref9}, as we take up our discussion of starting from the second pair than the one Dyson chose and then in seeking a variational principle in a minimal way, the other pair becomes a required ``adjoint" set, arriving inevitably at the familiar Lagrangian and its minus sign.

We will seek a variational construction of Maxwell's equations. First, in this and two succeeding paragraphs, we give a brief summary of the general procedure we adopt for constructing variational principles that considers the defining equations of the system as constraints in evaluating some desired property of the system \cite{ref10,ref11}. Thus, if

\begin{equation}
B_i(\psi) = 0,
\label{eqn1}
\end{equation}
are a set of equations defining a system characterized by functions $\psi$ (which may themselves be not a single but a set of functions), suppose we wish to evaluate some property $F(\psi)$ that depends on the solutions $\psi$. The notation of $\psi$ as possibly a wave function with the Schr\"{o}dinger equation and a normalization condition as two $B_i$ defining it is merely suggestive, our considerations not restricted to quantum physics. Of course, we generally confront a situation when the equations cannot be solved exactly for $\psi$ and the variational principle sought is an expression that yields the unknown $F(\psi)$ to second order errors, that is, if $\psi_t$ are some inexact, ``trial," solutions of $B_i(\psi) = 0$ that differ in first order,
   
\begin{equation}
\delta \psi  \equiv \psi_t -\psi,
\label{eqn2}
\end{equation}
this being a definition of $\delta \psi$, the desired variational expression $F_{v}$ must be such that $F_{v} -F(\psi)$ does not have terms of $O(\delta \psi)$ but only second- and higher-order terms.

The central feature of this construction \cite{ref10,ref11} is to incorporate the defining equations through Lagrange multipliers and write for the desired variational principle:

\begin{equation}
F_v = F(\psi_t) - L_{it}^\dagger B_i(\psi_t),
\label{eqn3}
\end{equation}
where we have introduced trial approximations to some unknown Lagrange multipliers $L_i$, as many as there are corresponding $B_i$, and summation over $i$ is presumed. At this point, we do not know what these exact $L_i$ are but Eq.~(\ref{eqn3}) has been written down to satisfy the obvious and minimal requirement that if, by some happy accident, the $\psi_t$ are exact solutions, then Eq.~(\ref{eqn3}) gives the exact value of $F$ desired, regardless of the value of the exact or trial Lagrange multipliers. Eq.~(\ref{eqn3}) is also broadly general, applicable to a variety of systems and quantities. Thus, if $F$ is a scalar and $B_i$ a vector quantity, then $L_i$ is also a vector and a dot product is implied. If $B_i$ is a functional of the functions $\psi$, then $L_i$ is also a functional such that the second term agrees with the functional form of $F$. The Lagrange multipliers, therefore, may be constants, functions of one or more variables, integral operators, all as required by the functional form of Eq.~(\ref{eqn3}) and $F$ and $B$ \cite{ref10,ref11}.

The next step of the construction is that with $\psi_t$ in general not expected to be exact, we ``make" the expression in Eq.~(\ref{eqn3}) variational, that is, we set all first-order terms on the right hand side equal to zero, a step that now defines what the Lagrange multipliers need to be. For this, as in Eq.~(\ref{eqn2}), we define first-order error in $L$ as 

\begin{equation}
\delta L \equiv L_t -L.
\label{eqn4}
\end{equation}
The coefficients of $\delta L$, being the defining equations in Eq.~(\ref{eqn1}), are zero by definition, and we now set the coefficient of $\delta \psi$ to zero to get the equations satisfied by the exact Lagrange ``functions." This may involve standard operations such as integration by parts if differentials have to be transferred to act on such $L$. In principle, it will give a set of well-defined equations for the exact Lagrange multipliers. The variational principle is now complete, approximate solutions of this set of ``adjoint" equations and of the original ones in Eq.~(\ref{eqn1}), when inserted into Eq.~(\ref{eqn3}), guaranteed by the very construction to contain no first-order errors in either the $\delta \psi$ or the $\delta L$ but only second-order errors. These include terms like $\delta L \delta \psi$ but note that the linearity of the introduction of $L$ in Eq.~(\ref{eqn3}) means that no higher order terms in $\delta L$ appear explicitly. In general, the adjoint Lagrange $L$ and the adjoint equations will differ from Eq.~(\ref{eqn1}), only special cases such as the famous Rayleigh-Ritz result for the ground state energy of a self-adjoint Hamiltonian having them coincide \cite{ref10,ref11}. Also, the sign of the second-order error in $F_v-F$ is unknown, the principle being stationary in that first-order departures vanish but nothing said about the sign of the second-order terms left behind. Only in special cases, such as again Rayleigh-Ritz, are they of well-defined sign in which case we have something even more powerful, a variational bound or extremum principle. 

To apply the above ideas to get a variational form of Maxwell's equations, we start with the two pairs as in Dyson's discussion \cite{ref2}

\begin{eqnarray}
\vec{\nabla} \cdot \vec{E} - \rho  & = & 0, \\ 
\frac{1}{c} \frac{\partial \vec{E}}{\partial t} -\vec{\nabla} \times \vec{B} +\frac{1}{c} \vec{j} & = & 0,
\label{eqn5}
\end{eqnarray}
and

\begin{eqnarray}
\vec{\nabla} \cdot \vec{B}  & = & 0, \\  
\frac{1}{c} \frac{\partial \vec{B}}{\partial t} -\vec{\nabla} \times \vec{E} & = & 0,
\label{eqn6}
\end{eqnarray}
with $\rho$ a charge and $\vec{j}$ a current density. As stated, we use Heaviside-Lorentz units \cite{ref8,ref12}.

We now seek a variational functional from which will follow the above Maxwell equations in as minimal a way as possible. For the choice of functional, a natural candidate is the energy of the field. If we view Eqs.(5-8) as defining equations Eq.~(\ref{eqn1}), with unknown $\psi$ being the electric and magnetic fields, and write a variational principle for the energy: 

\begin{equation}
\mathcal{E} = \frac{1}{2} \int dVdt [\vec{E}^2 + \vec{B}^2],
\label{eqn9}
\end{equation} 
with an integral over all volume and time, that would involve several Lagrange multipliers to incorporate the two vector and two scalar equations. Instead, by imposing a minimal requirement that no new functions or equations be introduced, let us write the form in Eq.~(\ref{eqn3}) as

\begin{equation}
\mathcal{E}_v = \frac{1}{2} \int dVdt [\vec{E}_t^2 + \vec{B}_t^2] - \int dVdt L_{1t}[\vec\nabla \cdot \vec{E}_t - \rho] -\int dVdt \vec{L}_{2t} \cdot [\frac{1}{c} \frac{\partial \vec{E}_t}{\partial t} -\vec{\nabla} \times \vec{B_t} +\frac{1}{c} \vec{j}],
\label{eqn10}
\end{equation}
where we have incorporated the first pair in Eqs.(5,6) with a scalar and a vector Lagrange multiplier as required to get an overall scalar expression for the energy. These Lagrange multipliers are to be determined. A comment about the integrations over space volume and time, the former of course natural to make an energy out of the energy density. The integration over time may appear strange but is done for the purposes of the variational formalism and getting differentials of both space and time transferred to the left upon integration by parts, and this requires also a time integral. This will appear more natural below when we are led to the Lagrangian, rather than the energy, density.  

Carrying out the construction of the variational principle by now writing in Eq.~(\ref{eqn10}), $\vec{E}_t = \vec{E} + \vec{\delta E}, \vec{B}_t = \vec{B} + \vec{\delta B}, L_{1t} = L_1 + \delta L_1, \vec{L}_{2t} = \vec{L}_2 + \vec{\delta L}_2$, and setting all first order coefficients to zero requires integration by parts to switch the $\vec\nabla$ and $\partial /\partial t$ to act on $L_1$ and $L_2$. A simple aspect of this integration by parts, but of profound consequence to what follows, is that such a switch of a first derivative as in time or of space when involved as a scalar product gives a minus sign whereas a switch of the cross product $\vec{\nabla} \times$ does not. Assuming that no surface terms arise, the $\vec{\delta E}$ and $\vec{\delta B}$ terms give, respectively,

\begin{eqnarray}
-\nabla L_1 -\frac{1}{c} \frac{\partial \vec{L}_2}{\partial t} & = & \vec{E} \\ \nonumber
\vec{\nabla} \times \vec{L}_2  & = & -\vec{B}.
\label{eqn11}
\end{eqnarray}

These equations are immediately suggestive to every physics student of the definitions of electric and magnetic fields in terms of scalar $\Phi$  and vector ($\vec{A}$) potentials were we to identify them with $L_1$ and $\vec{L}_2$, respectively, except for a crucial element, an extra minus sign in the second equation. The fix is also evident, to go back to Eq.~(\ref{eqn10}), and change the sign of the $\vec{B}_t^2$ in the first term on the right-hand side, with the attendant conclusion that what stands on the left-hand side should be not an integral over the energy density but rather the Lagrangian density, in which case this expression is one for the action. This also makes the inclusion of integration over time natural. Thus, these few steps have led inexorably to the Lagrangian while at the same time interpreting the Lagrange multipliers as the gauge potentials and, as is well known, that the definitions in Eq.~(\ref{eqn11}), with that minus sign dropped in the second equation, are equivalent to the second set of Maxwell's equations given in Eqs.(7,8). 

Thus, in a pithy fashion, we have one set of Maxwell's equations requiring the other as Lagrange or adjoint functions and leading naturally to the Lagrangian density and action thereof from which all four follow as the compatible equations of motion. There are no extraneous elements, no redundancy, only a tight and consistent set of arguments, all accessible even to a beginning student. We set out with two of Maxwell's equations, sought a variational form by a natural choice of energy but were led inexorably by their own internal logic to the gauge potentials of electromagnetism that are equivalent to the other pair of Maxwell's equations and to switching a sign so that it is not energy but the action that works for what we sought. Interestingly, \cite{ref7} stated that it ``does not make sense to assume one of Maxwell's equations to derive Maxwell's equations themselves."

And, by our very construction, when only approximate solutions of Maxwell's equations are available, calculating $\frac{1}{2} \int dVdt [\vec{E}_t^2 - \vec{B}_t^2]$ will have first-order errors for the action $\mathcal{A}$, but from Eq.~(\ref{eqn10}),

\begin{equation}
\mathcal{A}_v = \frac{1}{2} \int dVdt [\vec{E}_t^2 - \vec{B}_t^2] - \int dVdt L_{1t}[\vec\nabla \cdot \vec{E}_t - \rho] -\int dVdt \vec{L}_{2t} \cdot [\frac{1}{c} \frac{\partial \vec{E}_t}{\partial t} -\vec{\nabla} \times \vec{B_t} +\frac{1}{c} \vec{j}]
\label{eqn12}
\end{equation}
will be variationally accurate, that is, give the value of the action with only second-order error. Of course, as elsewhere with Lagrangian variational principles, the actual value of the action is often of no interest, even so-called null Lagrangians widely used in field theories. What counts is the expression itself and the equations of motion that it leads to.

The reverse procedure of ``deriving" Maxwell's equations starting from a Lagrangian density $\mathcal{L}$ is, of course, familiar to all students of physics in electromagnetism, whether classical or quantum. Using a four-vector notation as a compact alternative to the discussion so far \cite{ref1,ref12}, one starts with 

\begin{equation}
\mathcal{L} = -\frac{1}{4} F_{\mu\nu}F_{\mu\nu} +\frac{1}{c}j_{\mu}A_{\mu},
\label{eqn13}
\end{equation}
with $F_{\mu\nu} =\partial_{\mu}A_{\nu}-\partial_{\nu}A_{\mu}$. The Euler-Lagrange equations then give all Maxwell equations

\begin{equation}
\partial_{\nu}F_{\mu\nu} =\frac{1}{c}j_{\mu}.
\label{eqn14}
\end{equation}
In such a well-ploughed field as electromagnetism, one does not expect to find anything new in content, as indeed Feynman concluded, but what might be of interest is a new perspective that our analysis provides. In the conventional approach, both the gauge potentials $A_{\mu}$ and the electric and magnetic fields that constitute the electromagnetic tensor $F_{\mu\nu}$ are invoked at the start, and there is no a priori reason for the form in Eq.~(\ref{eqn13}) except that it leads to the desired Eq.~(\ref{eqn14}). By contrast, we started only with the electric and magnetic fields and two of Maxwell's equations and then sought a variational principle on minimal grounds of not introducing extraneous elements as far as possible. The variational formalism needs adjoint functions and in identifying them with the other two Maxwell equations, the gauge potentials did not need to be put in by hand but arise naturally, while at the same time the natural guess of starting with the expression for energy is ``corrected," also in the logic of the derivation itself to arrive at the form of the Lagrangian with that crucial minus sign. In that same logic also lies that while the fields are two vectors with six components, the form of one vector and one scalar equation that they obey suggests as adjoints a vector and a scalar object, namely the gauge potentials' four components.

\end{document}